\documentstyle[12pt,aasms4]{article}

\begin{document}

\title{A Stellar Population Gradient in VII Zw 403 - Implications for 
the Formation of Blue Compact Dwarf Galaxies
\footnotemark[1]}

\author{Regina E. Schulte-Ladbeck}
\affil{University of Pittsburgh, Pittsburgh, PA 15260, USA,
rsl@phyast.pitt.edu}
\authoremail{rsl@phyast.pitt.edu}
\author{Ulrich Hopp}
\affil{Universit\"{a}tssternwarte M\"{u}nchen, M\"{u}nchen, FRG, 
hopp@usm.uni-muenchen.de}
\author{Mary M. Crone}
\affil{Skidmore College, Saratoga Springs, NY 12866, USA, mcrone@skidmore.edu}
\author{Laura Greggio}
\affil{Osservatorio Astronomico di Bologna, Bologna, Italy, and
Universit\"{a}tssternwarte M\"{u}nchen, M\"{u}nchen, FRG, greggio@usm.uni-muenchen.de}

\footnotetext[1]{Based on observations made with the NASA/ESA Hubble Space Telescope 
obtained, and supported in part through grant number AR-06404.01-95A, from the 
Space Telescope Science Institute, which is operated by the Association of 
Universities for Research in Astronomy, Inc., under NASA contract NAS 5-26555.}

\begin{abstract}

We present evidence for the existence of an old stellar
halo in the Blue Compact Dwarf galaxy VII~Zw~403.
VII~Zw~403 is the first Blue Compact Dwarf galaxy for which a clear spatial 
segregation of the {\it resolved} stellar 
content into a ``core-halo" structure 
is detected. Multicolor HST/WFPC2 observations indicate that 
active star formation occurs in the central region, 
but is strikingly absent at large radii.
Instead, a globular-cluster-like red giant branch  
suggests the presence of an old ($>$~10~Gyr) and metal poor 
($<$[Fe/H]$>$=-1.92) stellar population in the halo. While the vast majority  
of Blue Compact Dwarf galaxies has been recognized to possess halos 
of red color in ground-based surface photometry, 
our observations of VII~Zw~403 establish for the first time a direct 
correspondence between a red halo color and the presence of 
old, red giant stars.
If the halos of Blue Compact Dwarf galaxies are all home to such
ancient stellar populations, then the fossil record 
conflicts with delayed-formation scenarios for dwarfs. 
\end{abstract}

\keywords{Galaxies: compact --- galaxies: dwarf --- galaxies: evolution --- 
galaxies: halos
--- galaxies: individual (VII~Zw~403 = UGC~6456) --- galaxies: stellar
content}

\section{Introduction}

The stellar content and evolutionary history of Blue Compact Dwarf (BCD)  
galaxies have been a puzzle for some time.  
The low metal abundances derived from the H-II regions of BCDs
(between about Z$_\odot$/3 and Z$_\odot$/50, Thuan et al. 1994), 
combined with their high rates of star formation and large H-I masses
(Thuan \& Martin 1981), have raised
the question of whether BCDs are truly young galaxies 
which formed their first stars recently, 
or old galaxies which occasionally light up with bright, compact
starburst regions (Searle \& Sargent 1972, Searle et al. 1973, 
Izotov \& Thuan 1999). In their original definition of a ``young" galaxy, 
Searle \& Sargent specified this is
a galaxy which formed {\it most} of its stars in recent times. 
In subsequent years the argument over BCD ages has evolved into a more
polarized one, because it is important to understand whether or not there
might exist any local galaxies which are only now forming their very 
first generation of stars out of pristine gas. Such BCDs 
that are making {\it all} of their stars in the on-going starburst
could be considered to be local examples of primeval galaxies
(Thuan \& Izotov 1998) and they would be fundamentally different from
old galaxies which began forming their first stars over 10~Gyrs ago at
high redshifts. 
 
Several lines of evidence support the old-galaxy hypothesis. 
Thuan (1983) found indications for the 
presence of evolved stars from the 
integrated, near-IR colors of BCDs. Fanelli et al. (1988) used 
UV spectra to show that the star-formation rates 
of BCDs are discontinuous. A deep CCD imaging survey by Loose \& Thuan (1986) 
first revealed that the bright, compact star-forming regions of BCDs
are embedded in much fainter, redder halos with elliptical 
outer isophotes. Nearly all galaxies in their sample ($>$95\%) show
such an underlying, low-surface-brightness component. 
Subsequent CCD surveys have confirmed this result 
(e.g., Kunth et al. 1988, Papaderos et al. 1996, 
Telles et al. 1997, Meurer 1999). Evolutionary synthesis
models (e.g., Kr\"{u}ger et al. 1991) and chemodynamical models 
(e.g., Rieschick \& Hensler 1998) favor the old-galaxy view as well. 

Hubble Space Telescope (HST) observations of several extremely metal-poor BCDs
recently led Thuan at al. (1997) and Thuan \& Izotov (1998) to propose that 
primeval BCDs may yet be found locally. Their argument 
is based in part on the blue colors observed for the unresolved underlying
disks. This is seen in I~Zw~18 (Z$_\odot$/50) by Hunter \& Thronson (1995), 
in SBS~0335-052 (Z$_\odot$/41) and SBS~0335-052W (around Z$_\odot$/50)
by Thuan at al. (1997), Papaderos et al. (1998) and Lipovetsky et al. (1999), 
and in SBS~1415+437 (Z$_\odot$/21) by Thuan et al. (1999). 
Based on abundance measurements, Izotov \& Thuan (1999) suggest that in fact 
{\it all galaxies with Z~$\le$~Z$_\odot$/20 are young}, with ages not 
exceeding 40~Myr, while those with
Z~$\le$~Z$_\odot$/5 are no older than $\approx$~1-2~Gyr. In the definition of
Izotov \& Thuan then, a ``young" galaxy is a galaxy which has not
experienced any star-forming events prior to the current one 
(which might have enriched its gas beyond that which is observed). 
By most accounts, even galaxies with ages $<$~2~Gyr would be 
considered to be relatively young galaxies, as their
formation redshift would not have been co-eval with that of
the large or other dwarf galaxies.

The controversy over the age of BCDs reflects our current knowledge 
of galaxy formation and evolution. 
Did dwarf galaxies form from primordial 
density fluctuations at z~$>>$~5 (Ikeuchi \& Norman 1987)?
Are the dwarfs observed today the leftover building blocks of  
large galaxies (White \& Rees 1978, Dekel \& Silk 1986)? 
Is the formation of dwarfs delayed until z~$\approx$~1 when
the UV background cooled
sufficiently for gas to collapse within 
small dark matter halos 
and form stars 
(Blanchard et al. 1992, Babul \& Rees 1992)?
Are dwarfs rapidly evolving for z~$<$~1, in which case they could
account for the rapid evolution in the galaxy luminosity function
necessary to explain the faint blue excess (Broadhurst et al. 1988, 
Babul \& Ferguson 1996, 
Spaans \& Norman 1997, Ferguson \& Babul 1998, Guzm\'{a}n et al. 1998)?
Where are the local, isolated H-I clouds still awaiting 
their first generation of stars (Briggs 1997)?

Resolved stellar populations are the fossil record
of a galaxy's star-formation history (SFH). They can also be used
to provide a definition of a galaxy's age. We shall consider ``young"
a galaxy which contains only ``young" stars, stars with ages
$<$~100~Myr. Intermediate-age stars signal a galaxy is at least
of intermediate age; we consider this to be the case when stars
that are older than a few hundred Myr are detected. Since the time-resolution of 
color-magnitude diagrams (CMD) of distant galaxies rapidly decreases if the
stellar ages exceed 1~Gyr, a detection of stars with
ages of about 1~Gyr is often considered sufficient to show
the presence of old stars. We, however, prefer to consider
all stars with ages of up to 10~Gyr as intermediate-age stars,
and reserve the term ``old" for stars of the kind that inhabit
Galactic globular clusters and have ages $>$~10~Gyr. An ``old"
galaxy is therefore one that has formed at least some stars
more than 10~Gyrs ago. 

The dwarf galaxies in the
Local Group exhibit an astounding variety of SFHs, yet all
contain old stars (Mateo 1998) in the above sense.
However, there are no BCDs known in the Local Group.
VII~Zw~403 is hence a key object; it is so close that HST
observations resolve it into 
individual stars (Schulte-Ladbeck et al. 1998, hereafter SCH98, 
Lynds et al. 1998). Thus it has become possible for the first time
to derive the distance of a BCD using a stellar distance
indicator (rather than just a recession velocity). This is crucial 
for the interpretation of the stellar content.

Recent determinations of the present-day metallicity of
the ionized gas in VII~Zw~403 (or rather, its O/H ratio),
were published by Martin (1997) and Izotov, Thuan \&  Lipovetsky (1997).
Martin finds log~(O/H) = -4.42(0.06), and Izotov, Thuan \&  Lipovetsky 
give -4.31(0.01). Both groups employ the metallicity scale for
which the Sun has log~(O/H) = -3.07. We note that Izotov \& Thuan 
(1999) also use this solar value. On this scale, the
metallicity of VII~Zw~403 is between 1/22 and 1/17 of solar. 

Despite its low metallicity ($\approx$~Z$_\odot$/20)
SCH98 argued that VII~Zw~403 is not a young galaxy, 
based on the detection of a red giant branch (RGB) 
with a well defined tip and a red asymptotic giant 
branch (AGB). VII~Zw~403 also
exhibits extended outer isophotes (Loose \& Thuan 1986, 
Hopp \& Schulte-Ladbeck 1995) with a red color that is
consistent with an old, metal poor population
(Schulte-Ladbeck \& Hopp 1998). The spectrum of the background
sheet displays absorption lines that indicate 
an evolved population underlying the recent starburst
(Hopp et al. 1998).   

In this paper, we provide further arguments that the existence of
an early epoch of
star-formation may be gleaned from the morphology of VII~Zw~403. We 
demonstrate that in radial bins outward from its starburst center,
the contribution to the CMD 
by young stars decreases, while
the red tangle that contains the old and metal-poor
RGB (Aparicio \& Gallart 1994) becomes a narrow feature. 
For radii larger than about 1~kpc, the young stars 
are absent and the stellar content is well described by a 
globular-cluster-like stellar population. 
We argue that this resolved old stellar population 
supports earlier suggestions
that the faint halos of BCDs harbor old stars.

\section{Observations and reductions}

HST WFPC2 observations of VII~Zw~403 were obtained in the continuum
(F336W, F555W and F814W filters, approximately the Johnson-Cousins U, V, and
I bands) and in the H$\alpha$ emission line (F656N filter). The relevant
exposures are listed in Table~1 of SCH98.

As described in SCH98, we re-ran the pipeline with improved
calibration files, removed cosmic rays, corrected for CTE, and 
corrected for geometric distortion. We conducted photometry on H$\alpha$-subtracted 
images.
We used DAOPHOT to perform psf photometry on each of the four WFPC2 chips. 
We calibrated the photometry using the most up-to-date SYNPHOT tables. 
After determining the positions of
each point source, we merged the
object catalogs of the four chips into a single file of
positions and instrumental magnitudes in the Vega magnitude system.
The accuracy of these coordinates relative to the guide star system
is about 0.5". 

In Figures 1a,b, we provide the errors for the point-source measurements
in the continuum filters, as well as the results of our completeness tests.
DAOPHOT/ALLSTAR residuals in F555W and F814W filters on all chips 
can be summarized by stating that they
reach 0.1~mag at magnitudes of about 26 
and 25.5, respectively. We checked the completeness of our 
photometry for each chip by adding a distribution of false stars 
consistent with the magnitude distribution of the real stars. The
percentage of recovered stars indicates that completeness is about
50\% at m$_{F555W}$~=~26, m$_{F814W}$~=~25. (Fig.~4 of SCH98 indicates
there are ``holes" in the distribution of red stars on the PC due to
incompleteness effects; this presumably also produces a less well populated
red tangle in the first two panels of Fig.~3 below.)

The foreground galactic reddening for VII~Zw~403 is E(B-V)~=~0.025
(Burstein \& Heiles 1984); we correct for the corresponding extinction
using the tables provided in Holtzman et al. (1995, Tables 12a,b).
A small and patchy internal reddening (E(B-V)$<$0.16) cannot be
ruled out (Lynds et al. 1998), but is not central to the arguments
presented below (the location of the blue plume in the CMD is consistent
with no or low internal reddening for the majority of the young stars
detected, the old stars in ``Baade's red sheet" are measured
throughout the halo where we find no evidence for internal extinction).

Final colors are transformed U, V, and I magnitudes using the 
color terms in Holtzman et al. (1995, Table 7).

\section{Results} 

Figure~2 is a position plot of all stars detected in both V and I (5459 objects).
We used the distribution of sources detected in both U and V to
locate the center of star fomation in VII~Zw~403 (R.A.[$^o$] = 172.002,
Dec.[$^o$] = 78.994, J2000). We then cast concentric circles about this 
location. In Figure~3, we display the [(V-I)$_o$, I$_o$] CMDs observed within each of
six radial bins marked in Fig.~2. The CMDs of the first and
sixth bins are also superimposed with stellar evolutionary tracks.
These tracks use the stellar atmospheres of F. Castelli (see, e.g.,
Bessell et al. 1998) for a metallicity of Z$_\odot$/30, which were
folded with the HST filter/system response and kindly made available to us by
L.~Origlia. Stellar evolution is based on the Fagotto et al. (1994) isochrones
for Z=0.0004 (Z$_\odot$/50). These models were chosen because they most
closely represent a ``compromise" metallicity for the stellar populations
which we observe (i.e., without trying to model the enrichment history, on which
we comment below, in detail as well). As is always the case, the comparison
of theoretical tracks (or synthetic CMDs) with data depends on the accuracy with which
we can ascertain either one and thus map one onto the other. 
We will consider as safe, conclusions which
do not depend on the choice of a particular model. For instance, a
well populated RGB is usually only observed if a stellar population with
an age upwards of about 1~Gyr is present, irrespective of the stellar metallicities adopted
(cf. also, Sweigart, Greggio \& Renzini 1990). We will point out where appropriate,
when inferences are based on what we consider more uncertain model results.
The location of the tracks on the observed CMDs employs the
distance modulus derived below.

The changes in stellar content with position are quite striking. 
The center displays a CMD that
has a dominant blue plume of main-sequence (MS) stars and 
blue supergiants (BSG) or blue-loop (BL) stars, a prominent red supergiant 
(RSG) plume, a few, very red AGB stars, and a weak red tangle. 
In the second bin, this red tangle and
the region of intermediate-mass BL stars are much more populated. 
We note that H$\alpha$ emission is detected only in
the first and second bins. 
By the third bin, both the blue plume and the
RSG plume have weakened whereas the red tangle is strong. 
By the fourth bin, most stellar indicators of
young ages have disappeared. Our detection 
limit of V$_o$~$\approx$~28 is low enough to show that  
MS stars younger than about 200~Myr are absent. 
The few remaining BL stars, which suggest ages of a few hundred Myr, 
are still present in this bin. In the fifth and sixth bins, we see
the outer regions of the galaxy. Here, the red tangle has become a
narrow band. 
There are seven objects that are red and brighter than I$_o$=23. 
A likely explanation is that these are Galactic foreground stars. 
The number of Galactic foreground stars within the WFPC2 
images is expected to be very small (M\'{e}ndez \& Guzm\'{a}n 1998);
since the fifth and sixth bins cover the largest area they
may contain a few. The CMDs of the fifth and
sixth bins suggest that only intermediate-age and 
old AGB and RGB stars are found in the outer regions of VII~Zw~403. 

In SCH98 we used the tip-of-the-red-giant-branch (TRGB) method to
obtain the distance to VII~Zw~403. 
Using the red tangle in the star-forming region,
we estimated a metallicity [Fe/H] of -1.2, 
resulting in M$_I$=-4.10 for the TRGB and 
a distance modulus of 28.4 mag  
(with a total random error of 0.09, and a dominant systematic 
error of 0.18).  This corresponds to a distance of 4.8~Mpc. 
We now re-derive the distance using the CMD of the outer
bins, only. The new I$_{TRGB}$, 24.25$\pm$0.05, is not significantly
different from our old value. Following Lee et al. (1993), we  
can estimate [Fe/H] from the V-I color either just below the TRGB, 
at M$_I$=-3.5, or better, at M$_I$=-3.0. We have done both and 
find a consistent (V-I)$_o$ color of 1.28, with an rms error 
of 0.03 (for M$_I$=-3.5) and
a dispersion of 0.21. This translates into a mean metallicity of
$<$[Fe/H]$>$=-1.92$\pm$0.04 (or Z=0.00024 or Z$_\odot$/83) with a spread or 
metallicity range (corrected for measurement error) of $\pm$0.7. 
We now find the TRGB at M$_I$=-3.98, or m-M=28.23, or a distance of 
about 4.4~Mpc, in excellent agreement with the value of 4.5~Mpc 
derived by Lynds et al. (1998). 

At this distance, 1" corresponds to about 21.5~pc. The
Holmberg diameter of VII~Zw~403 was measured by Schulte-Ladbeck
\& Hopp (1998) as 2a$_H$=146", and translates into a physical
diameter of 3.1~kpc. Fig.~3 shows that the young stars are 
contained within the inner few hundred pc, and
are absent beyond a radius of 1~kpc. Not surprisingly, this agrees
with the ``typical" size of a BCD (Thuan \& Martin 1981).

In Fig.~2 we colored in green the two radii which clearly separate
the young and old stellar components of VII~Zw~403. In Fig.~4,
we display the CMDs inside of the inner (386~pc) and outside 
of the outer (1352~pc) radius in blue and red, respectively. 
This illustrates the change in width of the red tangle due to
a diminishing contribution by intermediate-mass stars.
We mark the position of the TRGB, and provide an absolute-magnitude scale.
We also overlay  the empirical globular-cluster ridge-lines of
da Costa \& Armandroff (1990), which indicate that the halo population
of VII~Zw~403 is similar to the population of Galactic globular clusters, the
prototypes of Population~II. 

Fig.~4 shows that AGB stars, which populate a strip
from [(V-I)$_o$, M$_I$$_o$]$\approx$[1, -4] to [3.5, -6.5]
are found both at small and large radii. The strip may contain 
stars as old as 10~Gyr and Z=0.0004 (Z$_\odot$/50) at its bright, 
blue end. The extent of the AGB to the red requires 1/5$>$Z$_\odot$$>$1/20; here
the average location of stars is well described by the 4~Gyr
isochrones (Bertelli et al. 1994).   
Note that theoretical models of the AGB phase are very uncertain. This is
because a) their atmospheres contain difficult-to-model
molecules, and b) mass-loss is an important but ill-known parameter
that determines their evolution. If the AGB models can be believed, then
VII~Zw~403 has had an interesting chemical history.

Schulte-Ladbeck \& Hopp (1998) showed that their surface-brightness profiles
of VII~Zw~403 could be fit with an exponential law, with scale lenghts
in B of 25.7", R of 25.2". In Fig.~5, we display the I-band surface density 
of resolved stars versus radius, derived from starcounts in
the complete quadrant marked in Fig.~2 by blue lines.
An exponential law describes well the distribution of the resolved 
stars outside of the innermost region where there is incompleteness.
The scale length for all stars in I is 25.6" (or about 550~pc), in excellent 
agreement with our ground-based results. For the red 
stars (0.6$<$V-I$<$1.5) only, the scale length is 31.8" (or about 680~pc).

Owing to our observation that the CMD of the halo of VII~Zw~403 is
very reminiscent of that of a dwarf Spheroidal galaxy (dSph), and in connection 
with the interesting
question of what is the nature of the faint blue galaxies, we compare the
scale length of VII~Zw~403 to that of dSphs. We find that the scale length is
about 1.5 times larger than those of the largest Local Group 
dSph like NGC~147, NGC~185, NGC~205, or Fornax (Mateo 1998). 

We may ask how typical is this scale length for BCDs in general? One note
of caution --- other BCDs do not have direct distance determinations, and
the recession velocities in the literature are scaled with (different values of) Ho,
resulting in fairly unreliable scale lengths. Nevertheless, the distance errors
might average out over a large sample of scale lengths. We compared the scale 
length of VII~Zw~403 to values derived for the 
BCD-dominated samples of Bothun et al. (1989) and Telles et al. (1997). 
In both papers, deep CCD exposures to relatively large angular diameters were used for
detailed surface photometry and exponential scale lengths were
derived. The central starburst contribution was ignored in the determination. 
Bothun et al. used r-band data, Telles et al.,
V-band observations. We transformed both data sets to the same distance scale with
Ho = 75~km/s/Mpc. In the Bothun et al. sample, 24\% of the objects
have scale lengths of 0.7~kpc and smaller. Telles et al. found 38\% of
their objects in this regime.  Surface photometry of 93 galaxies from
the emission line galaxy sample of Popescu et al. (1997) was performed 
by Vennik, Hopp \& Popescu (1999). Again, this sample is dominated by
BCDs, but contains a few galaxies of other types. The scale lengths are
derived from R images, and 44\% of the galaxies show scale lenghts of 0.7~kpc 
and smaller. Papaderos et al. (1996) studied a sample of
14 galaxies in much more detail than the other authors. Here, the
derived R-band scale lengths range from 0.17 to 2.3~kpc. Despite the
various and different selection effects of the four samples from the
literature, we conclude from this comparison that the scale length
found for the underlying galaxy in VII~Zw~403 is fairly typical for
that of BCDs.

\section{Discussion}

Several pieces of evidence now suggest that VII~Zw~403 has an old
stellar halo, and this has implications for the formation of BCDs.
 
While it has previously been assumed that red colors at large radii
point to the presence of evolved stellar background sheets in BCDs, the
RSG, AGB and RGB stars actually overlap over a wide range in effective temperatures
and hence colors (cf. Fig.~6). Therefore, a red background sheet cannot be considered
as evidence of an old, underlying host galaxy unless it can be proven
that the stellar population which dominates the integrated color of a BCD is an 
old one. This has now been accomplished with the detection of red giants 
in the halo of VII~Zw~403.  

However, owing to the 
well-known age-metallicity degeneracy, comparatively young ($\approx$~1~Gyr) and 
metal rich RGB stars can populate the same region
of a CMD as old (10-15~Gyr) and metal poor RGB stars, 
based on optical, broad-band colors. 
We interpret the color dispersion of
the RGB with a variation in chemical abundance; however, age
dispersion (and/or differential reddening)
can also broaden the RGB (e.g., Bertelli et al. 1994).
In the case of VII~Zw~403 we argue that the well-defined RGB tip, 
and the narrowing of the red tangle from the center toward the 
outer regions of the galaxy where young stars are absent, 
indicate that our interpretation is appropriate. 
Conclusive evidence could be 
obtained from the detection of old horizontal-branch stars or the turn-off
of an old MS, but these are too faint to be reached with the HST. 
VII~Zw~403 does not appear to have a globular-cluster system.

Recently, the stellar halos of several dwarf Irregular (dIrr) 
and transition dIrr/dSph galaxies in the Local Group 
have been resolved into stars (WLM, Minniti \& Zijlstra 1996,
NGC~3109, Minniti et al. 1999, Antlia, Aparicio et al. 1997).  
These observations have been interpreted to indicate the 
presence of old and metal-poor, Population~II halos.
VII~Zw~403 joins the ranks of an increasing number of local, 
star-forming dwarf galaxies with resolved old stellar halos. 
As is pointed out by Mateo (1998), all
suitably studied star-forming dwarfs of the Local Group
show evidence for extended, smooth, and symmetric
distributions for their older stars.
Most BCDs show extended, smooth and red outer isophotes
as well. Our results for VII~Zw~403 imply the identification
of the background sheets with old stellar halos 
is justified for these BCDs.

The controversy over the age of BCDs ---  whether they are young or old
galaxies --- may be resolved in the following way: There is a continuum
of star-formation and chemical-enrichment histories among the BCDs. It
is likely that the vast majority of BCDs have an ancient ($>$~10~Gyr) stellar
population substratum, and must thus be recognized as old galaxies. This is
based on the observation that over 95\% of BCDs in the Loose \& Thuan sample
show extended background sheets of red color.
Comparing observations of the background-sheet colors 
for a sample of BCDs and dIrrs with the population 
synthesis models of Schmidt et al. (1985), Schulte-Ladbeck
\& Hopp (1998) suggest that complex star-formation histories 
prevail in these galaxies. Thus, depending on just how much mass is 
involved in the on-going starburst, and how the morphology of the 
star-forming regions compares to that of the older stellar substratum,
the colors of some BCDs might be entirely consistent with those of
``young" galaxies in the sense that they are currently experiencing a strong
starburst. If the outer isophotes are sufficiently red in color, 
they can be interpreted to indicate that such BCDs formed some stars at
epochs similar to that of Galactic globular cluster formation
(Kunth et al. 1988, Papaderos et al. 1996, Telles et al. 1997, Schulte-Ladbeck
\& Hopp 1998, Meurer 1999). The range of background-sheet colors suggests that
the SFHs of most BCDs since their formation at high redshift 
have probably been diverse, depending on the frequency, duration, and 
intensity of the star-formation events. This is similar to what has been
derived for dwarf galaxies within our Local Group. 

In a few of the extremely metal-poor BCDs, the color gradients
are small and the outer isophotes remain fairly blue 
(Hunter \& Thronson 1995, Thuan at al. 1997, Papaderos et al. 1998, 
Lipovetsky et al. 1999, Thuan et al. 1999).
The argument that such {\it blue} background-sheet colors indicate a 
galaxy is only now making its first generation of stars seems untenable to us in the light
of population synthesis models. Where the young and old stellar populations 
are spatially co-existent, Schmidt et al. (1985) show that a 
starburst completely dominates integrated optical colors for
up to 50~Myr, even if as little as 0.1\% of the dwarf galaxy's
mass is involved; hence a young burst may render the 
underlying old population undetectable. It is therefore possible
(although in the absence of deep CMDs not yet demonstrated) 
that the old populations in these few extremely metal-poor BCDs elude us 
due to a contrast problem.

On the other hand, the CMDs of the few BCDs which have been resolved with HST into
single stars indicate they do contain stellar generations which predate the
present starburst. In Figure 6, we show the evolution of
VII~Zw~403 in a series of synthetic ``snapshot" CMDs. These synthetic CMDs
were computed with the above Z$_\odot$/50 evolutionary
tracks. We used the Bologna code, which was recently adapted by Greggio et al.
(1998) for the simulation of HST data. The synthetic CMDs help to 
illustrate the general features of
stellar evolution at low metallicity, and may also serve as templates
for future CMDs of extremely metal-poor BCDs. The panels of Fig.~6 show
very well that the observed changes in stellar population with radius (Fig.~3)
can be interpreted as a change of the stellar ages with distance from the
core. We use synthetic CMDs to place a lower 
limit on the age of the red giants that we see in the halo of VII~Zw~403. 
We find a well defined TRGB first appears for ages $>$~3~Gyr (and
of course continues to be present up to 15~Gyrs). Beyond an
age of about 3~Gyr, we lose age resolution in the CMD; and we cannot
constrain, from the location of the RGB alone, the presence of stars with
ages in excess of 10~Gyr. Clearly, VII~Zw~403 has had a rich history of
star formation.

Aloisi et al. (1999) recently used synthetic CMDs to investigate deep 
HST CMDs of I~Zw~18, the most metal-poor BCD known (Z$_\odot$/50, from the
ionized gas). They
find that the present burst is not the first one to occur
in this galaxy either; the data require a prior burst 500~Myr to 1~Gyr ago. While
by most accounts a galaxy with an age below 1~Gyr would be considered
a young galaxy, these results for I~Zw~18 are in contradiction
with the primeval galaxy hypothesis of Izotov \& Thuan (1999) 
based on abundance analyses.

Unfortunately, there are no extremely metal-poor BCDs known
that are closeby enough to allow for look-back times of a large
fraction of the Hubble time. Whether or not we are observing
a very small percentage of extremely metal-poor BCDs while 
they are undergoing their very first starburst at the present epoch
is an interesting suggestion which remains to be investigated further; 
it will depend critically on the 
cycling of their gas (see below).

VII~Zw~403 shows evidence for at least three 
``eras" of star formation. The RGB stars suggest the first event
occured at a look-back time of at least 3~Gyr and probably $>$10~Gyr, 
the AGB stars indicate star-formation also happened at around 4~Gyr ago
(with considerable uncertainty), and 
the young stars testify to activity which took place less 
than about 1~Gyr ago. The data also allow us to infer 
the chemical enrichment history of VII~Zw~403. 
The RGB stars are consistent
with a metallicity of the order of Z$_\odot$/100,
while the ionized gas and (by assumption) the young stars
are at $\approx$Z$_\odot$/20 (Martin 1997, Izotov et al. 1997).
Curiously, an extended AGB is not expected unless
Z/Z$_\odot$$>$1/20 (see SCH98, Lynds et al. 1998).
The enrichment history of VII~Zw~403, at face value, 
is therefore inconsistent with closed-box models of galaxy evolution and  
seems to require the loss of enriched gas or accretion of metal-poor gas. 
VII~Zw~403 is isolated from massive neighbors, so an accrection
scenario \`{a} la Silk et al. (1987) seems unlikely. Papaderos et al. (1994) 
claim the X-ray detection of an outflow of hot gas from this galaxy. 
Employing our new distance, we can
estimate the total gas (H-I mass) from the measurements of 
Thuan \& Martin (1981) and Tully et al. (1981) to be about
7x10$^7$~M$_\odot$. 
This places VII~Zw~403 in the mass range for which models
suggest that gas may be blown out,
but the entire gas reservoir may not be blown away (Mac Low \& Ferrara 1999) .

\section{Conclusions}

The BCD VII~Zw~403 exhibits a radial population gradient. Young, blue stars
and H$\alpha$ emission are confined to the core. The core region has a 
diameter which equals the defining size of a BCD. Intermediate-age and old, 
red stars are distributed throughout an extended background sheet or halo.  
The halo stars show an RGB with a well
defined tip and are interpreted to be an old and metal-poor
population, similar to that in Galactic globular clusters.
BCDs were once recognized as ``the first metal-poor systems 
of Population I to be discovered'' (Searle \& Sargent 1972). 
VII~Zw~403 is the first BCD with compelling evidence for the existence
of a Population II halo. 
VII~Zw~403 is also the first BCD for which a direct comparison has been 
possible between the results from population synthesis of the
integrated halo color and resolved stellar content. The detection
of red giants at large distances from the starburst center 
verifies previous identifications of red halos with old stellar populations. 
If all BCD halos harbor 
old stars, then they must have formed at high redshift and survived re-heating; 
BCDs would not require the delayed-formation scenario.

\acknowledgments Work on this project was supported through an archival
research grant and guest observer HST grants to RSL and MMC (GO-7859.01-96A). 
UH acknowledges financial support from SFB375. We thank Livia Origlia
for supplying us with data prior to publication.

\clearpage

\figcaption[] {The photometric errors for the continuum filters (a). Notice
that in the F336W filter, most of the stars which are also found in F555W are
located on the PC chip; errors in the WF chips are
very large beyond a magnitude of 24 since many of the sources measured
are spurious detections. The fraction
of stars un-recovered in completeness tests (b).  Again, the results for 
the F336W filter on the WF chips show a much greater uncertainty because
those chips contain fewer real stars.}

\figcaption[]{A positional plot of all stars detected in V and I. Circles
drawn about the center have radii (in degrees) of 0.002, 0.005, 0.01, 0.0125,
0.0175, and 0.025, and are used to show the radial change of stellar content in
Figure~3. The green circles refer to the old versus young demarcation
of Figure~4. The blue lines mark the quadrant used for the stellar density 
profile in Figure~5.}

\figcaption[]{CMDs for each of the circular bins marked in Figure~2, 
with the radii translated from angular scale to physical scale.
Padua tracks with Z=0.0004 for stellar masses of 1, 4, and 15~M$_\odot$ 
are overlayed in green.}

\figcaption[]{CMD showing stars located in the core (blue) and those
found in the halo (red).  The dashed green line marks the TRGB, and
is the basis for the absolute-magnitude scale. The solid green lines are the
empirical globular cluster ridge-lines of da Costa and Armandroff (1990) for
(from left to right) [Fe/H] = -2.17, -1.91, -1.58, -1.54, -1.29, and -0.71. 
Error bars are given for the mean (V-I) color of the RGB. The slanted
shape of the red tangle is due to increasing errors and incompleteness
towards fainter magnitudes.}

\figcaption[]{The logarithm of the surface density of resolved stars
versus radius, with in black, all I detections, in red, only the red stars,
and structural parameters, $\alpha$$^{-1}$, for each.}

\figcaption[]{Synthetic CMDs providing snapshots of stellar evolution at
Z$_\odot$/50, computed for the distance, photometric errors, and completeness
numbers of VII~Zw~403. These CMDs serve as templates for the interpretation of
the observed CMDs. We used a ``compromise" metallicity, set between that of
the oldest and the youngest stars, although the chemical evolution of this
galaxy may not have been one of linear enrichment with time. The stars were 
randomly picked up from a grid in the mass range from 0.6 to 120~M$_\odot$
according to a standard Salpeter initial mass function; 
each panel displays 1000 surviving stars in the age interval indicated, while during 
the given age interval star-formation was assumed constant. We found it educational to
color-code different evolutionary phases: black -- main-sequence stars,
purple -- very massive post-MS stars, blue -- core-He burning (or blue-loop)
stars, green -- HRD crossing intermediate-mass stars, 
black -- RSG \& AGB stars, red -- RGB stars. In comparison with the data these template
CMDs illustrate the disappearance of the MS with age, the appearance of the RGB at
ages $>$~1~Gyr, and the sinking of metal-poor AGB stars below the TRGB for ages
$>$~3~Gry. The synthetic CMDs do not reproduce the very red AGB stars which we observe; 
these only appear in models of about a 3 to 10 times higher metallicity.}

\end{document}